\documentclass[reprint,amsmath,amssymb,aps,pre]{revtex4-2}
\usepackage{graphicx}
\usepackage{color}
\usepackage{amsmath}
\usepackage{amssymb}
\usepackage{amsthm}
\usepackage{extarrows}
\usepackage{booktabs}
\usepackage{float}
\usepackage[colorlinks,urlcolor=blue,linkcolor=blue,anchorcolor=blue,citecolor=blue]{hyperref}

\definecolor{DB}{RGB}{0, 24, 191}
\usepackage{mathptmx}

\begin{document}


\title{Boltzmann Distribution from Invariance of Coarse-Graining-Scale and Energy-Shift}

\author{Weicheng Fu$^{1,2,3}$}
\email{fuweicheng@tsnu.edu.cn}
\author{Yisen Wang$^{3}$}
\author{Yong Zhang$^{4,3}$}
\author{Hong Zhao$^{4,3}$}

\affiliation{
$^1$ Department of Physics, Tianshui Normal University, Tianshui 741001, Gansu, China\\
$^2$ Key Laboratory of Atomic and Molecular Physics $\&$ Functional Material of Gansu Province, College of Physics and Electronic Engineering, Northwest Normal University, Lanzhou 730070, China\\
$^3$ Lanzhou Center for Theoretical Physics, Lanzhou University, Lanzhou 730000, Gansu, China\\
$^4$ Department of Physics, Xiamen University, Xiamen 361005, Fujian, China
}

\date{\today }


\begin{abstract}

We present a concise derivation of the Boltzmann form for single-particle energy distributions in classical many-body Hamiltonian systems. The derivation relies on two physical facts: coarse-graining-scale invariance of the empirical distribution and invariance under a uniform shift of the energy zero. These conditions uniquely yield the Boltzmann factor, whose parameter is fixed by the mean energy per particle. For separable Hamiltonians, the equilibrium weight factorizes into kinetic and configurational contributions sharing the same parameter, identified from the kinetic part as the inverse kinetic temperature. The principle extends to any physical quantity with a stationary distribution and translational invariance. It is illustrated in a one-dimensional diatomic hard-core gas and a nonlinear lattice chain, where it predicts velocity, energy, spacing, collision-time, and pressure-dependent displacement distributions in agreement with simulations. The lattice model further shows how harmonic elasticity, anharmonic corrections, internal pressure, and thermal expansion emerge from the same exponential equilibrium weights. Finally, the relationships among different ensembles are briefly discussed.

\end{abstract}

\maketitle

\section{Introduction}

The Boltzmann distribution is one of the central results of equilibrium statistical physics \cite{khinchin1949mathematical}. It underlies the canonical ensemble, the Gibbs formulation of thermal equilibrium, and the Maxwellian velocity distribution as a special consequence of quadratic kinetic energy \cite{gibbs1902elementary}. In conventional treatments, the Boltzmann factor is derived from the equal a priori probability postulate applied to a subsystem weakly coupled to a large reservoir \cite{planck1932heat,landau1980statistical,PhysRevLett.96.050403}, from most-probable occupation-number arguments \cite{ehrenfest2002conceptual,ufniarz2008}, from Darwin--Fowler steepest-descent methods \cite{pathria2011statistical}, or from entropy-maximization and free-energy variational principles \cite{PhysRev.106.620,Shore1980-SHOADO-2,PhysRevE.81.051133}. Dynamical routes based on detailed balance, Markov processes, and Boltzmann's $H$ theorem provide complementary explanations for why this distribution is selected by relaxation processes \cite{Metropolis1953,boltzmann1964lectures,Brush2003KineticTheory}.

Beyond these textbook approaches, several alternative derivations and characterizations have been proposed \cite{Hosoya2015,Brandenburger2019,QAwang2002,e18050192,Sandomirskiy2025}, including operational constructions \cite{Hosoya2015}, axiomatic derivations \cite{Brandenburger2019}, invariance-based arguments \cite{QAwang2002,e18050192}, and structural uniqueness results for the Boltzmann family \cite{Sandomirskiy2025}. Taken together, these studies show that the exponential equilibrium weight can emerge from distinct physical and mathematical principles.

In this work, we present compact formulation of equilibrium distributions in classical many-body Hamiltonian systems. Since the empirical distribution of a continuous variable, such as energy, is practically obtained through binning, the functional form of an equilibrium distribution should be insensitive to the choice of binning scale, provided that the coarse graining is sufficiently fine. We refer to this property as coarse-graining-scale invariance. A second physical input follows from the arbitrariness of the energy zero in conservative dynamics. Adding a constant to the Hamiltonian does not affect the equations of motion and therefore cannot change the physical state. Consequently, the relative probabilities assigned to two energy bins may depend only on the energy difference between the bins, rather than on their absolute energy values. These two requirements lead to a functional equation for probability ratios, whose nontrivial solution is the Boltzmann weight. In this sense, coarse-graining-scale invariance ensures a stable distributional form, whereas energy-shift invariance fixes this form to be exponential. For a separable Hamiltonian, the resulting equilibrium weight factorizes into kinetic and configurational parts characterized by the same parameter $\beta$, which is identified from the kinetic contribution as the inverse temperature. The same argument can be extended to other stationary distributions associated with variables that exhibit translational invariance.

We demonstrate this principle in two one-dimensional (1D) systems. For a diatomic hard-core gas, the theory yields the Maxwellian velocity distribution, the single-particle energy statistics, the spacing law, and the collision-time interval distribution, all of which are verified by event-driven simulations. For a nonlinear chain with fixed-length constraints, we derive the velocity statistics together with a pressure-dependent law for particle displacements. In the harmonic limit, the theory reduces to linear elasticity, whereas in the anharmonic regime it provides exact series representations and asymptotic limits. In particular, cubic nonlinearity determines the internal pressure at fixed volume and the thermal expansion at zero pressure. Molecular dynamics simulations confirm these predictions and show explicitly how the constraints generate exponential equilibrium weights and how nonlinear asymmetry enters thermodynamic observables. In the following, we will first outline the general theoretical framework, then apply it to the two representative models and compare the analytical predictions with numerical simulations. We conclude with a summary and discussion of the formulation's implications.

\section{Core Idea}\label{sec2}

For a classical Hamiltonian particle system with total energy $E$, the accessible range of energy (nonnegative) of a single particle is $[0,E]$. To estimate its distribution, this interval is divided into $M$ bins, and the probability $p_i$ is obtained from counting statistics for the bin centered at $E_i$ $(i=1,\ldots,M;~E_i>0)$. In equilibrium, the functional form of the resulting distribution is numerically insensitive to the bin width, or equivalently to $M$, once the coarse graining is sufficiently fine. We refer to this property as \emph{coarse-graining-scale invariance}. Thus, one has
\begin{equation}
p_i=\frac{1}{Z(\beta,M)}f(E_i;\beta),
\qquad
Z(\beta,M)=\sum_{i=1}^{M} f(E_i;\beta),
\end{equation}
where the functional form of $f$ is independent of $M$, and $\beta$ is a parameter characterizing the distribution.

\section{Theoretical Analysis}\label{sec3}

The normalization and fixed mean-energy constraints are
\begin{equation}\label{eq-conditions}
\sum_{i=1}^{M} p_i =1,
\qquad
\sum_{i=1}^{M} p_i E_i = \bar{E},
\end{equation}
where $\bar{E}$ denotes the average single-particle energy.

A second physical input is \emph{the arbitrariness of the energy zero in a conservative system}. Since a uniform shift of all energies leaves the physics unchanged, probability ratios must be invariant under $E\to E+\epsilon$. Thus they can depend only on energy differences:
\begin{equation}
\frac{p_i}{p_j}
=
\frac{f(E_i;\beta)}{f(E_j;\beta)}
=
\frac{f(E_i+\epsilon;\beta)}{f(E_j+\epsilon;\beta)}
=
g(E_i-E_j),
\end{equation}
where $g$ is a positive function satisfying $g(0)=1$ and $g(-u)=1/g(u)$. For any three bins $i,j,k$, probability ratios satisfy the transitivity relation
\begin{equation}
\frac{p_i}{p_k}
=
\frac{p_i}{p_j}
\frac{p_j}{p_k}.
\end{equation}
Using the definition of $g$, this gives
\begin{equation}
g(E_i-E_k)
=
g(E_i-E_j)g(E_j-E_k).
\end{equation}
Equivalently,
\begin{equation}\label{eq-multiplicative-Cauchy}
g(u+v)=g(u)g(v),
\end{equation}
which is the multiplicative Cauchy equation \cite{aczel1966lectures}. Under the assumed continuity of $g$, its nontrivial solutions are
\begin{equation}
g(u)=e^{-\beta u},
\end{equation}
where $\beta$ is a real parameter. Therefore, the probabilities take the Boltzmann form
\begin{equation}\label{eq-B-dis}
p_i = C\, e^{-\beta E_i},
\qquad E_i \in (0,E],
\end{equation}
where the normalization constant $C$ is determined by
\begin{equation}
C = \frac{1}{\sum_{i=1}^{M} e^{-\beta E_i}}
\equiv \frac{1}{Z(\beta,M)}.
\end{equation}
Substituting Eq.~(\ref{eq-B-dis}) into Eq.~(\ref{eq-conditions}) yields
\begin{equation}\label{eq-barE}
\sum_{i=1}^{M} p_i E_i
=-\frac{\partial}{\partial \beta} \ln Z(\beta,M)
=\bar{E},
\end{equation}
which shows that the parameter $\beta$ is determined by $\bar{E}$.

Therefore, once a stationary single-particle energy distribution exists, the combination of coarse-graining-scale invariance and energy-shift invariance uniquely determines a Boltzmann-type distribution. More generally, any physical quantity with a stationary distribution and translational invariance admits a nontrivial equilibrium distribution of exponential form. The formulation will be illustrated with two examples: a 1D diatomic hard-core gas \cite{casati1986energy} and a nonlinear lattice chain. For later use, we specify the thermodynamic limit considered here. In one dimension, this limit is defined by $N,L\to\infty$ at fixed particle density $n=N/L$. The total energy is extensive, $E\to\infty$, while the energy per particle $E/N$ remains finite. The number of energy bins $M$ is arbitrary in principle, but must be large enough to avoid a trivial distribution in which all sampled energies fall into a single bin. For convenience, we take $M=N$ without loss of generality, so that, in the thermodynamic limit, discrete sums over bins can be replaced by integrals.

\section{Example 1: 1D diatomic hard-core gases}\label{sec4}

We consider a diatomic hard-core gas system consisting of $N$ hard spheres with alternating masses arranged in a straight line, $m_{2i-1}=m_1=m+\delta$ and $m_{2i}=m_2=m-\delta$, with $i=1,\ldots,N/2$ and $\delta\in(0,m)$.
Particles interact only through perfectly elastic collisions. The Hamiltonian is
\begin{equation}
H=\sum_{i=1}^{N}\varepsilon_i,\qquad \varepsilon_i=\frac{m_i v_i^2}{2}.
\end{equation}
Here $\varepsilon_i$ is the $i$th particle's kinetic energy.

In the thermodynamic limit, the original discrete summation in Eq. (\ref{eq-conditions}) can be replaced by an integral. Then the single-particle velocity distribution is obtained from the Boltzmann weight, i.e.,
\begin{equation}
\rho(v)=\frac{e^{-\beta m v^2/2}}{Z(\beta)} ,
\end{equation}
where the particle index has been omitted for brevity. The normalization and fixed   mean energy constraint give
\begin{equation}
Z(\beta)=\int_{-\infty}^{\infty} e^{-\beta m v^2/2}\mathrm dv
=\sqrt{\frac{2\pi}{m\beta}},
\end{equation}
and
\begin{equation}
\frac{\langle m v^2\rangle}{2}
=\int_{-\infty}^{\infty}\rho(v)\frac{m v^2}{2}\mathrm dv=\frac{1}{2\beta}
=\frac{E}{N}.
\end{equation}
Defining the kinetic temperature as
\begin{equation}
T=\langle m v^2\rangle=\frac{1}{\beta}=\frac{2E}{N},
\end{equation}
one obtains the Maxwellian velocity distribution
\begin{equation}
\rho(v)=\sqrt{\frac{m}{2\pi T}}
e^{-\frac{m v^2}{2T}}.
\end{equation}
The explicit form of the single-particle energy distribution can be obtained from the velocity distribution as follows:
\begin{equation}\label{eq-dis-e-gas}
\rho(\varepsilon)
=\frac{1}{\sqrt{\pi T\varepsilon}}
e^{-{\varepsilon}/{T}},
\qquad \varepsilon>0 .
\end{equation}
In equilibrium, the probability for a particle at rest is zero.  We therefore extend the definition to $\varepsilon=0$ by setting $\rho(0)=0$. Likewise, for finite $T$, the probability that the total energy is concentrated on a single particle vanishes in the thermodynamic limit.
Note that the thermodynamic temperature is identical with the kinetic temperature except for the presence of Boltzmann's constant.

Similarly, adding a constant to the distance between all particles does not affect the collision dynamics. That is, the interparticle distance possesses translational invariance, and consequently, it follows an exponential distribution.

Let the box length be $L$, and introduce two virtual boundary particles at $q_0=0$ and $q_{N+1}=L$. Defining the spacings
\begin{equation}
x_i=q_i-q_{i-1}>0,
\end{equation}
which satisfy
\begin{equation}
\sum_{i=1}^{N+1}x_i=q_{N+1}-q_0=L.
\end{equation}
Likewise, translational invariance implies an exponential distribution, i.e.,
\begin{equation}
\rho(x)=\frac{e^{-\beta x}}{Z(\beta)},\qquad x>0.
\end{equation}
Note that here $\beta>0$ is a parameter characterizing the spacing distribution and is no longer the kinetic temperature defined earlier. In the thermodynamic limit, one has
\begin{equation}
Z(\beta)=\int_0^\infty e^{-\beta x}\mathrm dx=\frac{1}{\beta},
\end{equation}
and
\begin{equation}
\langle x\rangle=\int_0^\infty x\rho(x)\mathrm dx=\frac{1}{\beta}=\ell,
\end{equation}
where $\ell=L/(N+1)$ is the mean spacing. Thus we obtain
\begin{equation}\label{eq-gas-distri-spacing}
\rho(x)=\ell^{-1}e^{-x/\ell}.
\end{equation}
The same result holds under periodic boundary conditions.

On the other hand, a collision between neighboring particles occurs if and only if their relative speed
\begin{equation}
\theta=v_i-v_{i+1}
\end{equation}
is positive. Estimating the collision-time interval by $\Delta t=x/\bar{\theta}$ gives
\begin{equation}\label{eq-dis-dt-gas}
\rho(\Delta t)
=\tau^{-1}
e^{-{\Delta t}/{\tau}},
\end{equation}
where $\tau=\ell/\bar{\theta}$ is the mean free time and $\bar{\theta}$ is the mean relative speed. Since the relative speed is Gaussian, that is
\begin{equation}
\rho(\theta)
=\sqrt{\frac{\mu}{2\pi T}}
e^{-\frac{\mu\theta^2}{2T}},
\end{equation}
with reduced mass
\begin{equation}
\mu=\frac{m_1m_2}{m_1+m_2}.
\end{equation}
One obtains
\begin{equation}\label{eq-tau-gas}
\tau
=\frac{\ell}{\bar{\theta}}
=
\frac{\ell}{\int_0^\infty \rho(\theta)\theta\mathrm d\theta}
=\ell\sqrt{\dfrac{2\pi\mu}{T}}=\ell
\sqrt{\dfrac{\pi(m^2-\delta^2)}{mT}} .
\end{equation}

\begin{figure}[t]
  \centering
  \includegraphics[width=1\columnwidth]{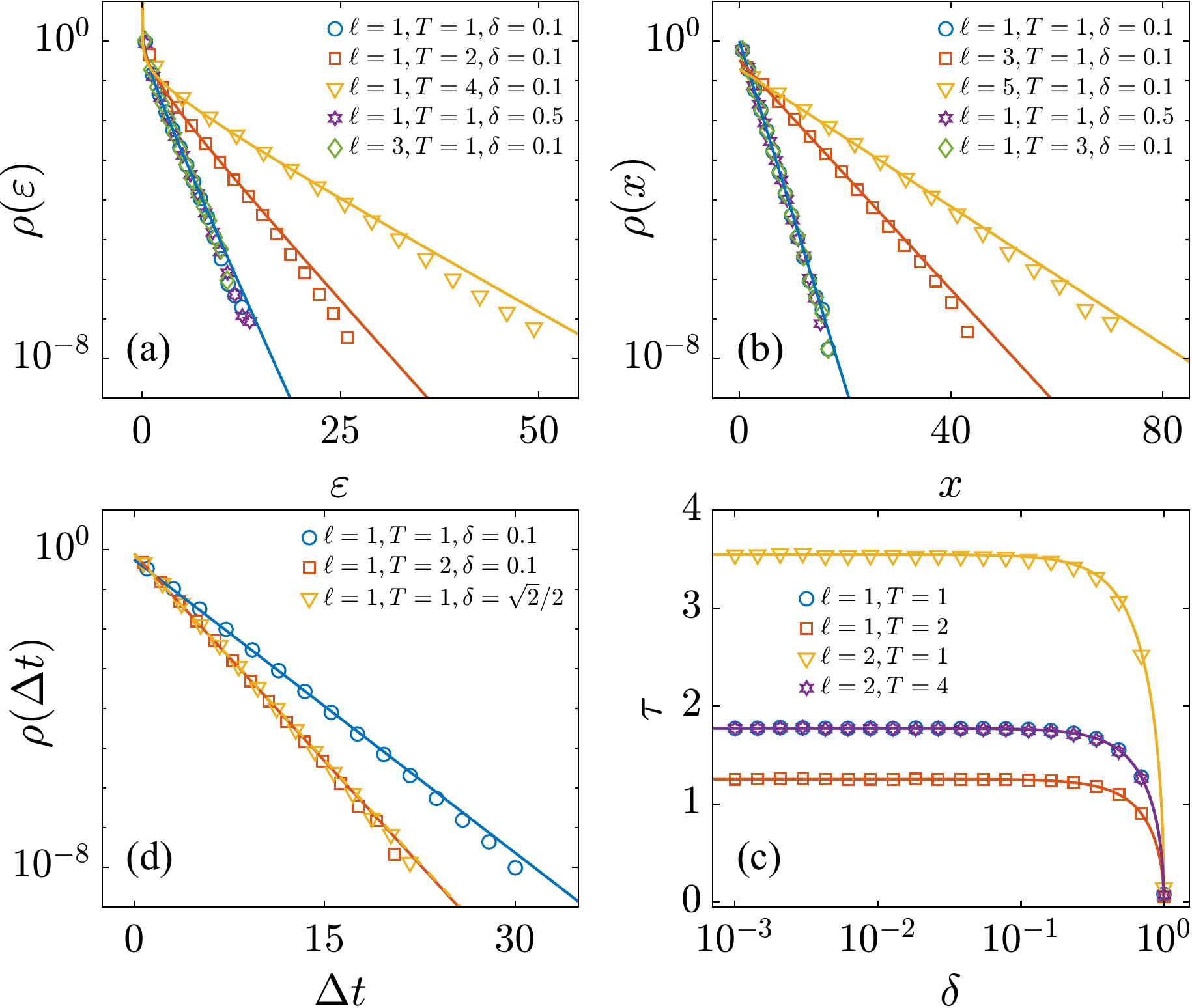}
  \caption{(a)--(c) Probability density functions of the particle energy $\rho(\varepsilon)$, interparticle spacing $\rho(x)$, and collision time interval $\rho(\Delta t)$ for different parameter sets. Symbols denote numerical results, while solid lines denote the corresponding theoretical predictions. (d) Mean free time $\tau$ as a function of the mass difference $\delta$. All results are obtained for $N=100$, $m=1$, and $\ell=1$.}\label{fig-gases}
\end{figure}

Figure~\ref{fig-gases} compares numerical simulations with theoretical predictions for different parameter settings at fixed system size $N=100$, $m=1$, and $\ell=1$. The system is updated using an event-driven algorithm \cite{PhysRevLett.89.180601}. As expected, the energy distribution $\rho(\varepsilon)$ depends only on the temperature $T$ [Eq. (\ref{eq-dis-e-gas})], while the spacing distribution $\rho(x)$ is controlled only by the mean spacing $\ell$ [Eq. (\ref{eq-gas-distri-spacing})]. In contrast, the collision-time interval distribution $\rho(\Delta t)$ depends on $T$, $\ell$, and the masses $m_i$ [Eqs. (\ref{eq-dis-dt-gas}) and (\ref{eq-tau-gas})]. The resulting dependence of the mean free time $\tau$ on $\delta$ is shown in Fig.~\ref{fig-gases}(d), see Eq. (\ref{eq-tau-gas}). All the numerical results (represented by different symbols) agree well  with the theoretical predictions (indicated by solid curves).

Note that, in the theoretical analysis, $M=N$ is taken. In the thermodynamic limit, the probabilities $p_i$ in Eq.~(\ref{eq-conditions}) become approximately continuous, allowing the discrete summation to be replaced by an integral and yielding a closed-form probability density function. In practice, by contrast, the probability density function is obtained by normalizing counts from discrete sampling, providing a discrete approximation. In experiments, $M$ does not need to be large when measuring the distribution of single-particle energy or interparticle spacing; Fig.~\ref{fig-gases} shows results for $M=15$. Changing $M$ affects only the density of data points, without altering the agreement between the numerical statistics and the theoretical curves (solid lines).

\section{Example 2: 1D lattice systems}\label{sec5}

We next consider a 1D nonlinear lattice composed of $N$ movable particles, whose Hamiltonian is
\begin{equation}\label{eq-Ei-lattice}
H=\sum_{i=1}^{N}\varepsilon_i,
\qquad
\varepsilon_i=
\frac{m_i v_i^2}{2}
+
V(q_{i+1}-q_i-a),
\end{equation}
where $\varepsilon_i$, $m_i$, $v_i$, and $q_i$ denote the energy, mass, velocity, and position of the $i$th particle, respectively. The parameter $a$ is the equilibrium lattice spacing in the absence of external strain. The nearest-neighbor interaction is taken as
\begin{equation}\label{eq-Vx-lattice}
V(x)=\frac{x^2}{2}+\frac{\alpha x^3}{3}+\frac{\lambda x^4}{4}.
\end{equation}

The boundary conditions determine the constraint on the interparticle spacing. For free boundaries, the length of chain is allowed to fluctuate. For periodic or fixed boundaries, the total length is fixed. Specifically, periodic boundaries satisfy $q_{N+1}=q_1+L$, while fixed boundaries can be represented by two immobile virtual particles at $q_0=0$ and $q_{N+1}=L$. Defining
\begin{equation}
\xi_i=q_{i+1}-q_i>0
\end{equation}
as the nearest-neighbor spacing, the mean spacing is $b=L/N$ for periodic boundaries and $b=L/(N+1)$ for fixed boundaries. The corresponding strain is
\begin{equation}
l=b-a,
\end{equation}
where $l>0$ and $l<0$ correspond to stretched and compressed chains, respectively.

From the Boltzmann form in Eq.~(\ref{eq-B-dis}), the single-particle energy distribution is
\begin{equation}\label{eq-lattice-distri-Ei}
\rho(\varepsilon_i)
=
\frac{e^{-\beta\varepsilon_i}}{Z(\beta)}.
\end{equation}
For a separable kinetic and configurational energy, the joint distribution of velocity and spacing factorizes as
\begin{equation}\label{eq-lattice-distri-xv}
\rho(\xi,v)
=
\frac{1}{Z(\beta)}
e^{-\beta\left[V(\xi-a)+\frac{m v^2}{2}\right]}
=
\rho(\xi)\rho(v),
\end{equation}
with
\begin{equation}\label{eq-lattice-distri-f}
\rho(\xi)
=
\frac{e^{-\beta V(\xi-a)}}{Z_{\xi}(\beta)}
\end{equation}
and
\begin{equation}\label{eq-lattice-distri-g}
\rho(v)
=
\frac{e^{-\beta m v^2/2}}{Z_v(\beta)}.
\end{equation}
Considering the velocity part, we have
\begin{equation}\label{eq-lattice-gv1}
Z_v(\beta)
=\int_{-\infty}^{\infty}
e^{-\frac{\beta m v^2}{2}}\mathrm dv
=\sqrt{\frac{2\pi}{m\beta}},
\end{equation}
and
\begin{equation}\label{eq-lattice-gv2}
\frac{\langle m v^2\rangle}{2}=\int_{-\infty}^{\infty}
\rho(v)\frac{m v^2}{2}\mathrm dv
=\frac{1}{2\beta}.
\end{equation}
Similarly, defining the kinetic temperature as
\begin{equation}
T=\langle m v^2\rangle=\frac{1}{\beta},
\end{equation}
one obtains the Maxwellian velocity distribution
\begin{equation}\label{eq-lattice-distri-g-end}
\rho(v)=\sqrt{\frac{m}{2\pi T}}e^{-\frac{m v^2}{2T}}.
\end{equation}

For free boundary conditions, the spacing variable $\xi$ is constrained only by the interaction potential, and its distribution is given by Eq.~(\ref{eq-lattice-distri-f}). This case should be distinguished from the spacing distribution in the diatomic hard-core gas. In the lattice systems, $\xi$ contributes to the potential energy and is therefore subject to the energy constraint. Consequently, as shown in Eq.~(\ref{eq-lattice-distri-xv}), the kinetic and configurational weights share the same parameter $\beta=1/T$. By contrast, in the hard-core gas, the interparticle spacing carries no potential energy and is thus independent of energy.

For periodic or fixed boundary conditions, the fixed total length imposes an additional constraint on the mean spacing. Consequently, the distribution of $\xi$ necessarily includes a constraint term analogous to Eq.~(\ref{eq-gas-distri-spacing}). In lattice systems, this constraint results in an additional weight
\begin{equation}\label{eq-lattice-distri-spacing}
h(\xi)\propto e^{-\beta_L \xi}.
\end{equation}
The spacing variable $\xi$ is therefore constrained both by the interaction potential and by the fixed-length condition. Multiplying the configurational Boltzmann weight by the constraint weight and renormalizing gives
\begin{equation}\label{eq-lattice-distri-spacing2}
\begin{aligned}
\tilde{\rho}(\xi)
&=
\frac{1}{Z_\xi(\beta,\beta_L)}
e^{-\beta V(\xi-a)-\beta_L \xi}  \\
&\equiv
\frac{1}{Z(T,P)}
e^{-\frac{V(\xi-a)+P\xi}{T}},
\end{aligned}
\end{equation}
with
\begin{equation}
Z(T,P) \equiv \displaystyle\int_{0}^{\infty} e^{-\frac{V(\xi-a)+P\xi}{T}}\mathrm d\xi,
\end{equation}
where $T=1/\beta$ and $P=\beta_L/\beta$ is a parameter determined by the imposed length constraint. In thermodynamic terms, $P$ actually refers to the internal pressure \cite{Spohn2014} that satisfies
\begin{equation}\label{eq-define-pressure}
P=-\left\langle \frac{\partial V}{\partial \xi}\right\rangle_{P,T}.
\end{equation}
For brevity, the tilde on $\rho$ will be omitted hereafter.

The imposed mean spacing is fixed by
\begin{equation}\label{eq-lattice-spacing-mean}
\int_{0}^{\infty}\rho(\xi)\xi\mathrm d\xi
=\langle \xi\rangle_{P,T}=b,
\end{equation}
which serves as the equation of state of the chain, connecting $T$, $P$, and the system length, with $L=Nb$ for periodic boundaries and $L=(N+1)b$ for fixed boundaries. The corresponding mean potential energy is
\begin{equation}\label{eq-lattice-spacing-ZbetaX}
\int_{0}^{\infty}
\rho(\xi)V(\xi-a)\,\mathrm d\xi
=
\langle V(\xi-a)\rangle_{P,T}.
\end{equation}

Introducing the relative displacement $x=\xi-a$,
one has
\begin{equation}\label{eq-lattice-spacing-meanX}
\langle x\rangle_{P,T}
=
\int_{-\infty}^{\infty}\rho(x)x\mathrm dx
=b-a=l,
\end{equation}
and
\begin{equation}\label{eq-lattice-spacing-meanVx}
\langle V(x)\rangle_{P,T}
=\int_{-\infty}^{\infty}\rho(x)V(x)\mathrm dx,
\end{equation}
where
\begin{equation}
\rho(x)
=\frac{1}{\widetilde Z(T,P)}
e^{-\frac{V(x)+Px}{T}},
\end{equation}
with
\begin{equation}
\widetilde Z(T,P)
=\int_{-\infty}^{\infty}
e^{-\frac{V(x)+Px}{T}}\mathrm dx .
\end{equation}
Strictly, the lower integration limit is $x=-a$, but extending it to $-\infty$ is an accurate and convenient approximation when the probability of classical particles passing through each other is negligible.

Interestingly, for free boundary conditions, the length constraint is absent, corresponding to $P=0$, so that Eq.~(\ref{eq-lattice-distri-spacing2}) reduces to Eq.~(\ref{eq-lattice-distri-f}). While, in the absence of interparticle interactions, Eq.~(\ref{eq-lattice-distri-spacing2}) reduces to the result for the 1D diatomic hard-core gas, Eq.~(\ref{eq-gas-distri-spacing}). Comparing the corresponding coefficients gives $P\ell=T$, consistent with the equation of state of 1D ideal gases \cite{boltzmann1964lectures}.

\subsection{The harmonic case}

For $\alpha=\lambda=0$, the model reduces to a harmonic chain. In this case,
\begin{equation}
\begin{cases}
Z(T,P)=\sqrt{2\pi T}\,
\exp\left[\dfrac{P(P-2a)}{2T}\right], \\[6pt]
P=-\langle x\rangle=a-b=-l, \\[6pt]
E=N\left(T+\dfrac{P^2}{2}\right)
=N\left[T+V(l)\right],
\end{cases}
\end{equation}
where $E$ is the total energy and $V(l)=l^2/2$ is the elastic energy associated with the imposed strain.

\subsection{The anharmonic cases}

For $\lambda<0$, the integral in Eq.~(\ref{eq-lattice-spacing-meanX}) is in general divergent. We therefore restrict ourselves to the stable case $\lambda>0$, for which the integral is well defined. Although no elementary closed form is available in general, the relevant quantities can be represented by a convergent series:
\begin{equation}
\begin{cases}
Z(T,P)=D\exp\left(-C/T\right),\\[6pt]
\langle x\rangle_{P,T}=-\dfrac{\alpha}{3\lambda}
+\dfrac{Y_1}{D},\\[6pt]
\langle V(x)\rangle_{P,T}
=\dfrac{\lambda}{4}\dfrac{Y_4}{D}
+A\dfrac{Y_2}{D}
+(B-P)\dfrac{Y_1}{D}
+C+\dfrac{P\alpha}{3\lambda},
\end{cases}
\end{equation}
where
\begin{equation*}
\begin{cases}
A=\dfrac{1}{2}-\dfrac{\alpha^2}{6\lambda},\\[6pt]
B=P-\dfrac{\alpha}{3\lambda}
+\dfrac{2\alpha^3}{27\lambda^2},\\[6pt]
C=-\dfrac{P\alpha}{3\lambda}
+\dfrac{\alpha^2}{18\lambda^2}
-\dfrac{\alpha^4}{108\lambda^3},\\[6pt]
D=\displaystyle\sum_{n=0}^{\infty}
\frac{1}{(2n)!}
\left(\frac{B}{T}\right)^{2n}
M_{2n},\\[6pt]
Y_1=-\displaystyle\sum_{n=0}^{\infty}
\frac{1}{(2n+1)!}
\left(\frac{B}{T}\right)^{2n+1}
M_{2n+2},\\[6pt]
Y_2=\displaystyle\sum_{n=0}^{\infty}
\frac{1}{(2n)!}
\left(\frac{B}{T}\right)^{2n}
M_{2n+2},\\[6pt]
Y_4=\displaystyle\sum_{n=0}^{\infty}
\frac{1}{(2n)!}
\left(\frac{B}{T}\right)^{2n}
M_{2n+4},\\[6pt]
M_{2n}
=e^{\frac{A^2}{2\lambda T}}
\left(\frac{2T}{\lambda}\right)^{\frac{2n+1}{4}}
\Gamma\left(n+\frac12\right)
D_{-(n+\frac12)}
\left(
\frac{A}{\sqrt{\lambda T/2}}
\right).
\end{cases}
\end{equation*}
Here $D_\nu(z)$ denotes the parabolic cylinder function \cite{abramowitz1972handbook}, which may be written in terms of the Kummer confluent hypergeometric function as
\begin{equation*}
\begin{aligned}
D_{\nu}(z)
=2^{\nu/2}e^{-z^2/4}\sqrt{\pi}
\Bigg[&
\frac{{}_1F_1\left(-\frac{\nu}{2};\frac12;\frac{z^2}{2}\right)
}{\Gamma\left(\frac{1-\nu}{2}\right)}
\\
&-\frac{\sqrt{2}z\,{}_1F_1\left(\frac{1-\nu}{2};\frac32;\frac{z^2}{2}\right)
}{\Gamma\left(-\frac{\nu}{2}\right)}
\Bigg],
\end{aligned}
\end{equation*}
where ${}_1F_1(a;b;z)$ is the Kummer function \cite{yang2021introduction}.

This exact representation is mathematically useful but does not provide a transparent physical picture. We therefore turn to two limiting regimes: the low-temperature limit and the high-temperature limit.

\subsubsection{Low-temperature limit}

We define the effective potential
\begin{equation}
U(x)=\frac{x^2}{2}+\frac{\alpha x^3}{3}
+\frac{\lambda x^4}{4}+Px .
\end{equation}
For sufficiently small $T$, the integral is dominated by the global minimum $x_*$ of $U(x)$, determined by
\begin{equation}
U'(x_*)=0,
\end{equation}
or explicitly
\begin{equation}\label{eq-cubic-equation}
x_*+\alpha x_*^2+\lambda x_*^3+P=0 .
\end{equation}
The corresponding stability condition is
\begin{equation}
U''(x_*)=1+2\alpha x_*+3\lambda x_*^2>0 .
\end{equation}
If $x_*$ is a nondegenerate global minimum, the Laplace expansion gives
\begin{equation}
\langle x\rangle
=
x_*-\frac{T}{2}
\frac{U'''(x_*)}{[U''(x_*)]^2}
+O(T^2).
\end{equation}
Using $U''(x)=1+2\alpha x+3\lambda x^2$ and $U'''(x)=2\alpha+6\lambda x$, we obtain
\begin{equation}\label{eq-mean-X-x-star}
\langle x\rangle
\simeq
x_*-
T
\frac{\alpha+3\lambda x_*}
{(1+2\alpha x_*+3\lambda x_*^2)^2}.
\end{equation}
Thus, to leading order, the mean relative displacement is fixed by the stable equilibrium of the effective potential,
\begin{equation}\label{eq-low-T-x-xstar}
\langle x\rangle\simeq x_*,
\end{equation}
while the first thermal correction is controlled by the local asymmetry of $U(x)$.

For weak pressure, the minimum $x_*$ can be expanded around the $P=0$ solution. Since $x_*=0$ at $P=0$, one finds
\begin{equation}
x_*=-P-\alpha P^2+O(P^3).
\end{equation}
Keeping the leading terms in both $P$ and $T$ yields
\begin{equation}\label{eq-meanX-P-T}
\langle x\rangle
\simeq
-P-\alpha T .
\end{equation}
The first term is the elastic displacement induced by the pressure field, whereas the second is the thermal bias generated by the cubic asymmetry of the potential.

For zero pressure, Eq.~(\ref{eq-meanX-P-T}) reduces to
\begin{equation}\label{eq-mean-X-x-star-P=0}
\langle x\rangle\simeq -\alpha T .
\end{equation}
Conversely, imposing zero mean strain, $\langle x\rangle=0$, gives
\begin{equation}\label{eq-mean-X-P-star-l=0}
P(T)\simeq -\alpha T .
\end{equation}

\subsubsection{High-temperature limit}

In the high-temperature limit, $T\to\infty$, with $\alpha$, $\lambda$, and $P$ fixed, the mean relative displacement has the asymptotic form
\begin{equation}\label{eq-meanX-high-T}
\langle x\rangle
=-\frac{\alpha}{3\lambda}
-2\frac{\Gamma(3/4)}{\Gamma(1/4)}
\frac{P-\frac{\alpha}{3\lambda}
+\frac{2\alpha^3}{27\lambda^2}}
{\sqrt{\lambda T}}+O(T^{-1}).
\end{equation}
Therefore, one has
\begin{equation}\label{eq-high-T-mean-X}
\lim_{T\to\infty}\langle x\rangle
=
-\frac{\alpha}{3\lambda},
\end{equation}
which indicates that in high-temperature limits, the relative displacement is entirely determined by the ratio of the cubic and quartic nonlinear coefficients.

If the mean strain is constrained to vanish, $\langle x\rangle=0$, Eq.~(\ref{eq-meanX-high-T}) gives
\begin{equation}
P(T)
=-\frac{\alpha\sqrt{\lambda T}}{6\lambda}
\frac{\Gamma(1/4)}{\Gamma(3/4)}+\frac{\alpha}{3\lambda}
-\frac{2\alpha^3}{27\lambda^2}+O(T^{-1/2}).
\end{equation}
The leading term is
\begin{equation}\label{eq-P-T-High-T}
P(T)\simeq-\frac{\alpha}{6\sqrt{\lambda}}
\frac{\Gamma(1/4)}{\Gamma(3/4)}\sqrt{T}
\approx-\frac{\alpha}{2\sqrt{\lambda}}\sqrt{T}.
\end{equation}
Hence, at high temperature, a fixed zero-strain condition cannot be maintained by a temperature-independent pressure. Instead, the required pressure scales as
\begin{equation}\label{eq-high-T-P}
P\propto -\alpha\sqrt{T/\lambda}.
\end{equation}

\subsection{Numerical simulations}

We now test the theoretical predictions for the nonlinear lattice chain by molecular dynamics simulations. In numerical simulations, a trick can be adopted to directly initialize the system into the target state.
For a prescribed target state $(N,P,T)$, Eq.~(\ref{eq-lattice-spacing-meanX}) determines the equilibrium spacing, $b=a+l$.
The particles are initially placed at
\begin{equation}
q_i=i\,b ,\quad i=1,\dots,N,
\end{equation}
so that the uniformly strained configuration defines the zero point of potential energy. The corresponding total energy is
\begin{equation}\label{eq-Chain-En}
E(N,P,T)=
N\left[{T}/{2}+\langle V(x)\rangle_{P,T}-V(l)\right].
\end{equation}
Here $T/2$ is the mean kinetic energy per particle, while $\langle V(x)\rangle_{P,T}-V(l)$ is the mean excess potential energy relative to the initial uniformly strained state. The initial velocities are sampled from the Maxwellian velocity distribution and then rescaled so that the total kinetic energy exactly equals $E(N,P,T)$. Specifically, we set
\begin{equation}
v_i\rightarrow \eta v_i,
\qquad
\eta=
\left(
\frac{2E(N,P,T)}
{\sum_{i=1}^{N} m_i v_i^2}
\right)^{1/2}.
\end{equation}
The system is then evolved under Hamiltonian dynamics until equilibrium is reached. In our simulations, the periodic boundary conditions are imposed, and the equations of motion are integrated using the eighth-order Yoshida symplectic integrator~\cite{YOSHIDA1990262}. The resulting dynamics is microcanonical: the total energy is conserved, whereas kinetic and potential energies are continuously exchanged. For an equilibrated trajectory, however, the relevant long-time averages are well defined. We define the kinetic temperature as $T=\langle m v^2\rangle$ and the internal pressure $P$ as the mean one-sided force, as specified in Eq.~(\ref{eq-define-pressure}). In experiment, all physical quantities are nondimensionalized. Without loss of generality, the particle masses and the lattice constant are set to unity, i.e., $m_i=1$ and $a=1$.

Within the present microcanonical simulations, $T$ and $P$ are not externally imposed reservoir parameters. Instead, they are determined self-consistently by the conserved total energy and fixed system length. They characterize the equilibrium distributions of particle velocities and relative displacements, and thereby determine the corresponding average observables. This interpretation is consistent with the usual thermodynamic meaning of temperature and pressure. Macroscopically, temperature characterizes the thermal state of a system, while microscopically it is related to the average kinetic energy of random particle motion. Similarly, the pressure appearing here is the mechanical force conjugate to the length constraint. Thus, although the simulations are performed in the $NVE$ ensemble, the parameters $T$ and $P$ play the same statistical role as temperature and pressure in the corresponding equilibrium distribution. In particular, the pressure appearing here is the mechanical force conjugate to the length constraint.

\begin{figure}[t]
  \centering
  \includegraphics[width=1\columnwidth]{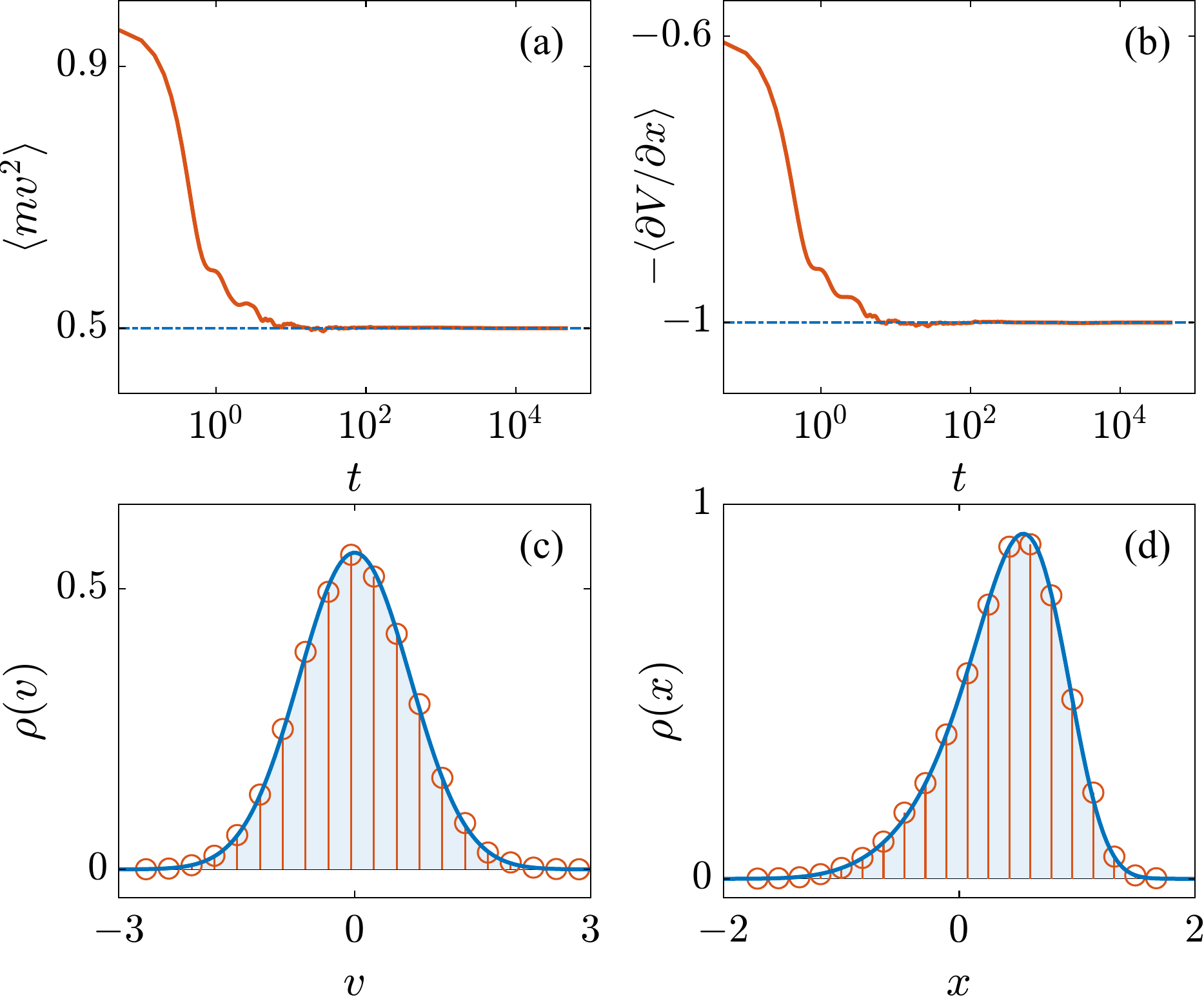}\\
  \caption{Molecular dynamics results for a lattice chain with periodic boundaries, $\alpha=\lambda=1$, $N=100$, $T=0.5$, and $P=-1$. (a, b) Time evolution of the average kinetic energy and internal pressure, respectively, showing rapid convergence to the theoretical values (see dashed lines). (c, d) Probability density functions of particle velocity and relative displacement $x$, respectively, compared with the corresponding theoretical curves.}
  \label{distribution_graph}
\end{figure}

\begin{figure*}[t]
  \centering
  \includegraphics[width=2\columnwidth]{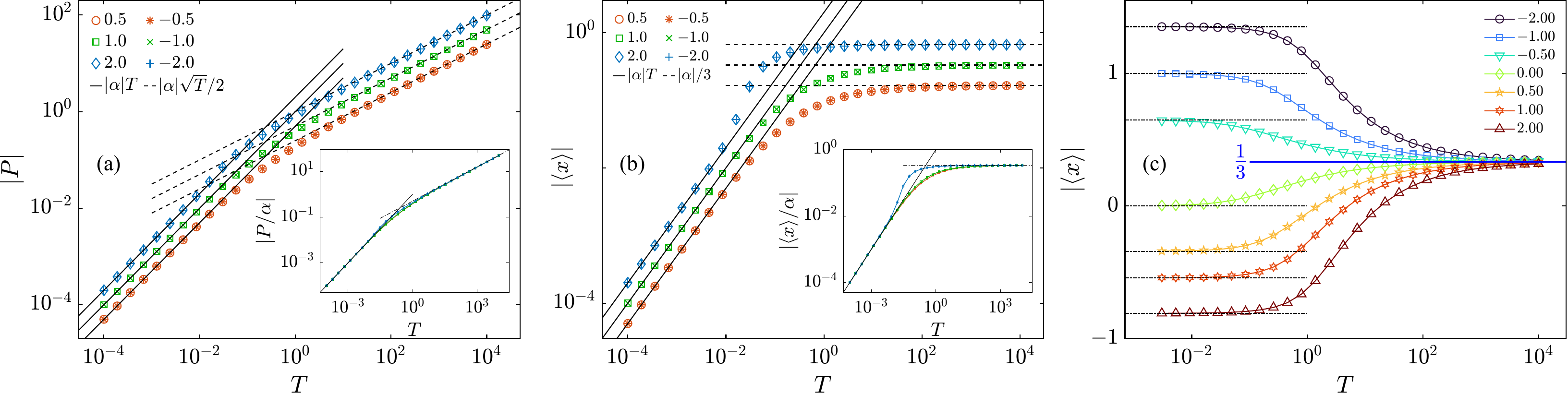}
  \caption{Temperature dependence of the pressure and mean displacement obtained from Eq.~(\ref{eq-lattice-spacing-meanX}). (a) $|P|$ versus $T$ at $\lambda=1$ and $l=0$ for different $\alpha$. (b) $|\langle x\rangle|$ versus $T$ at $\lambda=1$ and $P=0$ for different $\alpha$. (c) $|\langle x\rangle|$ versus $T$ at $\lambda=1$ and $\alpha=-1$ for different $P$. Symbols denote numerical solutions. Solid and dashed lines in (a) and (b) indicate the low- and high-temperature asymptotic predictions, respectively. In (c), the black dashed line marks the stable root $x_*$ of Eq.~(\ref{eq-cubic-equation}), and the blue solid line gives the high-temperature limit. Insets in (a) and (b) show the corresponding rescaled data.}\label{fig_PvsT}
\end{figure*}

Figure~\ref{distribution_graph} shows the simulation results for the lattice chain with $\alpha = \lambda = 1$, $N=100$, temperature $T=0.5$, and pressure $P=-1$. Under these settings, the corresponding $l \simeq 0.38848$, and $\langle V(x)\rangle_{P,T}\simeq 0.33363$, thus $E\simeq48.29329$. Panels (a) and (b) display the time evolution of the average kinetic energy and internal pressure, respectively, which rapidly converge to the theoretical values prescribed by the target parameters (see horizontal dotted lines). Panels (c) and (d) compare the numerical probability density functions of particle velocity $v$ and relative displacement $x$ with the corresponding theoretical curves (solid lines). The agreement between simulation and theory is excellent.

Figure~\ref{fig_PvsT} shows numerical solutions of Eq.~(\ref{eq-lattice-spacing-meanX}) for various system parameters:

(a) Absolute pressure $|P|$ versus temperature $T$ at $\lambda=1$ and $l=0$ for several values of $\alpha$ (see legend). Symbols indicate numerical results, while solid and dashed lines correspond to the low- and high-temperature theoretical limits, respectively [Eqs.~(\ref{eq-mean-X-P-star-l=0}) and (\ref{eq-P-T-High-T})]. The vertical axis of the main panel has been rescaled by $\alpha$ for clarity (see inset). The scaled curves collapse in both limits, with small deviations observed for $10^{-2}<T<7$.

(b) Absolute mean displacement $|\langle x\rangle|$ versus $T$ at $\lambda=1$ and $P=0$ for different $\alpha$ (see legend). Different symbols denote numerical results. Solid and dashed lines indicate the low- and high-temperature limits, respectively [Eqs.~(\ref{eq-mean-X-x-star-P=0}) and (\ref{eq-high-T-mean-X})].  After rescaling the vertical axis by $\alpha$ (see inset), the curves overlap in both limits, with some differences for $10^{-2}<T<75$.

(c) Absolute mean displacement $|\langle x\rangle|$ versus $T$ at $\lambda=1$ and $\alpha=-1$ for several $P$ values (see legend). Symbols represent numerical results. The black dashed line marks the real root $x_*$ of Eq.~(\ref{eq-cubic-equation}), which approximates $\langle x\rangle$ in the low-temperature limit [Eq.~(\ref{eq-low-T-x-xstar})]. The blue solid line shows the high-temperature theoretical value [Eq.~(\ref{eq-high-T-mean-X})], illustrating that at high $T$, $\langle x\rangle$ is dominated by the ratio of the cubic and quartic nonlinear coefficients in the interaction potential.

These results demonstrate that the third-order nonlinearity plays a central role: it sets the internal pressure under fixed-volume conditions and governs thermal expansion under zero-pressure conditions. Beyond its decisive influence on thermodynamic properties, it is also essential for transport~\cite{PhysRevE.85.060102,PhysRevE.88.052112,PhysRevE.89.032102} and relaxation phenomena, for example, by inducing anomalous phonon damping~\cite{Feng2022} and influencing the thermalization of nonlinear chains~\cite{Fu_2019}.

\section{Summary and discussion}\label{sec6}

In summary, we have presented a concise derivation of the Boltzmann distribution in classical Hamiltonian many-body systems. The derivation is based on two ingredients: coarse-graining-scale invariance of the empirical distribution and invariance under a uniform shift of the energy zero in conservative systems. These conditions lead directly to a functional equation whose nontrivial continuous solution is exponential, with the parameter determined by the mean single-particle energy. More generally, the same argument applies to any physical quantity that possesses a stationary distribution and translational invariance.

The framework was illustrated both theoretically and numerically using two 1D systems. These examples demonstrate that the formulation provides a compact and physically transparent approach for identifying how macroscopic constraints determine exponential equilibrium weights and how nonlinear asymmetry enters thermodynamic observables. Importantly, the analytical framework is independent of the system's spatial dimensionality: the quantity subject to binning statistics is scalar and thus does not involve spatial coordinates, and the translational invariance of the zero of potential energy is a global property that similarly does not depend on dimension.

Clearly, the exponential distribution is a one-parameter function, with the parameter fixed uniquely by the mean value of the statistic. For a classical system with separable Hamiltonian, $H=K(v)+V(q)$,
the equilibrium weight factorizes into kinetic and configurational parts, $e^{-\beta H}=e^{-\beta K(v)}e^{-\beta V(q)}$,
so that both parts share the same parameter $\beta$. Only consider the kinetic part, $\beta$ is naturally identified with the inverse kinetic temperature. In the $NVE$ ensemble, $T$ and $P$ are therefore not externally imposed control parameters, but derived thermodynamic quantities whose meanings are fixed by the microscopic averages defining the kinetic temperature and internal pressure. Similarly, in the $NPT$ ensemble, the energy and volume fluctuate, while their mean values are determined by the isothermal-isobaric distribution.
In the thermodynamic limit, the relative fluctuations of thermodynamic quantities decay scaling as $1/\sqrt{N}$ and converge to zero \cite{landau1980statistical}, and all statistical ensembles tend to be equivalent. Variables that are fixed in one ensemble may fluctuate in another, but the macroscopic constraint relations and the associated equilibrium distribution remain unchanged. The distinction between ensembles is therefore mainly reflected in which variables are controlled and which are allowed to fluctuate.

Moreover, the analysis presented in this paper assumes that the relevant statistical quantities, such as the single-particle energy, possess well-defined stationary distributions. In other words, the variables under consideration are assumed to sample all values allowed by the macroscopic constraints, so that the form of the distribution can be determined from the corresponding invariance properties of the system. This framework concerns the structure of equilibrium distributions rather than the detailed dynamical mechanism by which equilibrium is reached. If the system exhibits localization, nonergodicity, or other forms of restricted phase-space exploration, this assumption may fail and a separate dynamical analysis is required. Only when the dynamics is sufficiently mixing, so that memory of the initial condition is effectively lost and the long-time evolution explores the accessible region of phase space, can the system relax to equilibrium and be described by the Boltzmann distribution \cite{krylov1979foundations}. The functional equation used in this work also has a natural probabilistic interpretation [see again Eq. (\ref{eq-multiplicative-Cauchy})]. It is analogous to the multiplicative property of characteristic functions for independent random variables: factorization reflects statistical independence, while the corresponding nontrivial continuous solution has an exponential form. This provides another way to understand why the Boltzmann weight emerges from the combined requirements of stability under coarse graining and invariance under shifts of the energy zero.

\begin{acknowledgments}
This work was supported by the National Natural Science Foundation of China (Grants Nos. 12247106, 12465010, 12575040, 12575042). W. Fu also acknowledges support from the Long-yuan Youth Talents Project of Gansu Province, the Fei-tian Scholars Project of Gansu Province, the Leading Talent Project of Tianshui City, the Innovation Fund from the Department of Education of Gansu Province (Grant No.~2023A-106), the Project of Open Competition for the Best Candidates from Department of Education of Gansu Province (Grant No. 2021jyjbgs-06), and the Open Project Program of Key Laboratory of Atomic and Molecular Physics $\&$ Functional Material of Gansu Province (6016-202404).
\end{acknowledgments}


%

\end{document}